\documentclass[11pt,letterpaper]{article}
\pdfoutput=1
\usepackage{jcappub}
\usepackage{subcaption}
\usepackage{bbm,bm}
\usepackage{mathrsfs}
\usepackage{slashed}
\usepackage{caption}
\usepackage{epstopdf}
\usepackage[normalem]{ulem}
\usepackage[bottom]{footmisc}
\usepackage{subcaption}
\usepackage{bbold}
\usepackage{titlesec}
\usepackage{threeparttable}
\usepackage{booktabs}
\usepackage{changepage}
\usepackage[utf8]{inputenc}
\usepackage{soul}

\usepackage{grffile}

\usepackage{graphicx}  
\usepackage{dcolumn}   
\usepackage{bm}        
\usepackage{amssymb}   
\usepackage{setspace}
\usepackage{amsmath, amssymb, setspace}
\usepackage{array}
\usepackage{booktabs}
\usepackage{caption}
\usepackage{indentfirst}
\usepackage{float}
\usepackage{lmodern}
\usepackage{multirow}
\usepackage{soul}
\usepackage[normalem]{ulem}

\usepackage{braket}
\usepackage{comment}

\usepackage[draft]{pgf}

\newcommand{\DoBox}[1]{\begin{center}
\color{red}\fbox{
\begin{minipage}{0.9\textwidth}

\end{minipage}}
\end{center}}

\newlength{\myimageoversize}
\newsavebox{\myimage}

\titleformat{\subsubsection}
 {\normalfont\fontsize{12}{17}\itshape}{\thesubsubsection}{1em}{}

\title{\huge{Primordial black hole dark matter from catastrogenesis with unstable pseudo-Goldstone bosons}}

\author[a]{Graciela B. Gelmini,}
\author[a]{Jonah Hyman,}
\author[a]{Anna Simpson,}
\author[b]{and Edoardo Vitagliano}

\affiliation[a]{Department of Physics and Astronomy, University of California, Los Angeles\\
475 Portola Plaza, Los Angeles, CA 90095-1547, USA}

\affiliation[b]{Racah Institute of Physics, Hebrew University of Jerusalem, Jerusalem 91904, Israel}

\emailAdd{gelmini@physics.ucla.edu}
\emailAdd{jthyman@physics.ucla.edu}
\emailAdd{ansimps@g.ucla.edu}
\emailAdd{edoardo.vitagliano@mail.huji.ac.il}

\begin{document}

\abstract{ We propose a new scenario for the formation of asteroid-mass primordial black holes (PBHs). Our mechanism is based on the annihilation of the string-wall network associated with the breaking of a $U(1)$ global symmetry into a discrete $Z_N$ symmetry.  If the potential has multiple local minima ($N>1$) the network is stable, and the annihilation is guaranteed by a bias among the different vacua.  The collapse of the string-wall network is accompanied by \textit{catastrogenesis}, a large production of pseudo-Goldstone bosons (pGBs)---e.g. axions, ALPs, or majorons---gravitational waves, and PBHs. If pGBs rapidly decay into products that thermalize, as predicted e.g. in the high-quality QCD axion and heavy majoron models, they do not contribute to the dark matter population, but we show that PBHs can constitute 100\% of the dark matter. The gravitational wave background produced by catastrogenesis with heavy unstable axions, ALPs, or majorons could be visible in future interferometers. 
}

\maketitle

\section{Introduction}
\label{Intro}
The only known interaction that dark matter (DM) has with ordinary matter is gravitational~\cite{Bertone:2004pz,deSwart:2017heh}. 
This is why black holes can be
good DM candidates if they are formed in the early Universe, i.e. are primordial black holes (PBHs)~\cite{1967SvA....10..602Z,Hawking:1971ei,Carr:1974nx,Carr:1975qj,Carr:2020gox,Carr:2021bzv}.
Moreover, they need to be stable, so that they do not evaporate through Hawking's radiation before the present time~\cite{Carr:2009jm,Hawking:1974rv,Hawking:1975vcx, Boudaud:2018hqb,DeRocco:2019fjq,Coogan:2020tuf}, and they need to
avoid current constraints from microlensing~\cite{EROS-2:2006ryy,Niikura:2017zjd,Niikura:2019kqi}, as well as many other bounds collected in e.g. Ref.~\cite{PBHbounds}.
There is a window of 
PBH masses, between $10^{-16}$ and $10^{-10}\, M_\odot$ (roughly equivalent to asteroid masses), for which PBHs can constitute 100\% of DM. A necessarily incomplete list of proposed formation scenarios for PBHs in the asteroid-mass range includes density perturbations in the early Universe~(see e.g.~\cite{Carr:1975qj,Yokoyama:1995ex,Garcia-Bellido:1996mdl,Ballesteros:2017fsr}), bubble collisions~\cite{Hawking:1982ga,Lewicki:2019gmv}, the collapse of cosmic strings~\cite{Hawking:1987bn}, 
scalar field dynamics~\cite{Khlopov:1985jw,Cotner:2016cvr,Cotner:2018vug,Cotner:2019ykd}, 
long-range interactions~\cite{Flores:2020drq}, and collapse of domain walls~\cite{Garriga:2015fdk,Deng:2016vzb} or vacuum bubbles~\cite{Deng:2017uwc,Kusenko:2020pcg} in multi-field inflationary scenarios. We propose here a novel production scenario for asteroid-mass PBHs based on the mechanism we have dubbed catastrogenesis in our previous work~\cite{Gelmini:2022nim}. Our mechanism does not depend on inflationary physics, nor on primordial density perturbations, and crucially it can be linked to other unsolved problems in particle physics, e.g. the strong CP problem. 

We assume the existence of a global $U(1)$ symmetry whose spontaneous breaking is associated with the existence of Goldstone bosons. A generic feature of these models is that the global symmetry is also broken explicitly, so that the bosons are massive, i.e. pseudo-Goldstone bosons (pGBs). This happens in a plethora of extensions of the Standard Model of particle physics, including the original axion
model~\cite{Peccei:1977hh,Weinberg:1977ma,Wilczek:1977pj}, invisible axion (also called QCD axion) models~\cite{Kim:1979if,Shifman:1979if,Dine:1981rt}, generic axion-like particle (ALP) models~\cite{Jaeckel:2010ni}, and (singlet) majoron models~\cite{Chikashige:1980ui,Rothstein:1992rh,Gu:2010ys,Lazarides:2018aev,Reig:2019sok,Abe:2020dut,Bansal:2022zpi}. Intriguingly, some models predict a larger-than-expected mass for the QCD axion~\cite{Fukuda:2015ana, Dimopoulos:2016lvn,Agrawal:2017ksf,Gaillard:2018xgk}, including the so-called ``high-quality QCD axion''~\cite{Hook:2019qoh} and previous models. (For a review on earlier attempts of model building predicting heavy QCD axions, see Section 6.7 of Ref.~\cite{DiLuzio:2020wdo}.) Heavy majorons, even of mass in the TeV range, have been considered as well (see e.g.~\cite{Gu:2010ys,Abe:2020dut}). If the pGBs are unstable and decay in the early Universe, they could not constitute the DM, and thus such models might lack a DM candidate. This can be the case for axions and ALPs, as well as for majorons which could get a mass from soft breaking terms or from gravitational effects (see e.g.~\cite{Rothstein:1992rh,Gu:2010ys,Lazarides:2018aev,Reig:2019sok,Abe:2020dut}).

The stability of these pGBs depends on their mass and couplings. Axion-like particles with a coupling to photons are probed by beam-dump experiments~\cite{CHARM:1985anb,Riordan:1987aw,Dolan:2017osp,Blumlein:1990ay,NA64:2020qwq}, supernovae~\cite{Jaeckel:2017tud,Hoof:2022xbe,Lucente:2020whw,Caputo:2021rux,Caputo:2022mah,Caputo:2022rca,Diamond:2023scc} and neutron star mergers~\cite{Diamond:2023cto} up to several hundreds MeV, while cosmological observations~\cite{Cadamuro:2011fd,Depta:2020zbh} reach even larger masses, above a TeV, for very small couplings~\cite{Cadamuro:2011fd,Depta:2020zbh}.
 Colliders can exclude masses up to the TeV scale~\cite{Knapen:2016moh,Bauer:2017ris}, though only for large couplings (this implies e.g. an open window for the high-quality QCD axion~\cite{Hook:2019qoh}). Assuming couplings to leptons, there are bounds from beam dumps~\cite{Bauer:2017ris} and astrophysical observations~\cite{Caputo:2021rux,Ferreira:2022xlw}, but the parameter space is largely unconstrained above the GeV scale.\footnote{A collection of bounds on axions and axion-like particles can be found in Ref.~\cite{AxionLimits}.} Heavy majorons are similarly probed by cosmology~\cite{Kelly:2020aks},  supernovae~\cite{Fiorillo:2022cdq}, and laboratory searches~\cite{Berryman:2018ogk,deGouvea:2019qaz,Brdar:2020nbj}, which again leave masses above 1 GeV largely unconstrained.

We deal with bosons that are heavier that a GeV and decay very early into Standard Model products which rapidly thermalize. Axions would decay into photons and charged fermions. Majorons could decay mostly into two relatively heavy right-handed neutrinos, which in turn decay very fast into pions, charged fermion pairs, and active neutrinos. Therefore, such pGBs cannot be the DM in the models we consider, but catastrogenesis~\cite{Gelmini:2022nim} may lead to a different DM candidate, namely PBHs. We consider explicit breaking potentials which admit a number $N>1$ of minima along the orbit of vacua. In this case,
if inflation happens before the spontaneous symmetry breaking, as we assume, a cosmological problem  may arise:  a stable string-wall network forms and could eventually come to dominate the Universe leading to an unacceptable cosmology~\cite{Zeldovich:1974uw, Vilenkin:1984ib}. Both axion~\cite{Sikivie:1982qv} and majoron~\cite{Lazarides:2018aev} models can result in $N>1$.

The cosmological problem is solved by adding an additional explicit breaking  in the Lagrangian~\cite{Sikivie:1982qv,Chang:1998bq,Gelmini:1988sf}, possibly due to the effect of Planck suppressed operators~\cite{Barr:1992qq,Kamionkowski:1992mf,Gelmini:2022nim}, which produces an energy difference (a bias) between the minima that leads to a unique vacuum. As first proposed in 1974 in Ref.~\cite{Zeldovich:1974uw}, this small bias results in the annihilation of the string-wall network, and only one vacuum is eventually left. 
The annihilation is accompanied by catastrogenesis~\cite{Gelmini:2022nim}, i.e. the production of gravitational waves (GWs), pGBs, and PBHs. Our main result is that for a range of the string-wall network annihilation temperatures, one can form asteroid-mass PBHs and potentially a GW signal in future observatories. 
As we will see, both the abundance of PBHs and the amplitude of the stochastic GW background produced depend crucially on the details of the annihilation.

Although we focus on models featuring pGBs (that we will call, as customary in the recent literature, ALPs), our mechanism also applies in principle to the breaking of any discrete symmetry with multiple vacua, when the wall system annihilation produces unstable particles, such as flavons~\cite{Gelmini:2020bqg}, whose decay products thermalize.

The paper is structured as follows. In Sec.~\ref{Model} we introduce the generic particle model we assume and describe its cosmology. In Sec.~\ref{PBH} we present our estimates for the PBH relic abundance and find the range of parameters in which they could account for the whole of the DM. In Sec.~\ref{GW} we show that for the same range of parameters a signal in GWs could be observable in future detectors. Sec.~\ref{sec:consistency} explains in detail restrictions imposed by consistency conditions of our model. Finally the last section contains some concluding remarks.

 \section{ALP models and their cosmology}
 \label{Model}

Here we briefly review well-known aspects of the models we study. A more detailed account can be found e.g. in Ref.~\cite{Gelmini:2022nim} (see also~\cite{Gelmini:2021yzu}). We consider a complex scalar field $\phi= |\phi | e^{i \theta}$ whose phase after spontaneous symmetry  breaking  at an energy scale $V$ is related to the ALP field $a$ as $a = \theta\, V$. The generic potential $V(\phi)$  for the scalar field includes three terms, one with a spontaneously broken global $U(1)$  symmetry and two others which break the symmetry explicitly,  
\begin{align}
\label{eq:potential}
V(\phi) \supset~ &  \frac{\lambda}{4} (|\phi |^2- V^2)^2 + \frac{v^4}{2} \left(1- \frac{|\phi |}{V} \cos(N \theta) \right) 
 -  \epsilon_b  v^4 \frac{|\phi |}{V}  \cos \left(\theta - \delta\right),
\end{align}
 with $V \gg v$.   The first term is $U(1)$ invariant and leads to the spontaneous breaking of this symmetry during a phase transition in the early Universe, at which the magnitude of the field becomes $|\phi |\simeq V$ and cosmic strings are produced due to the random distribution of the phase $\theta$ in different correlation volumes.  Assuming that the bosons have the same average energy of the visible sector (which is possible in many inflation models), this phase transition happens at a temperature $T\simeq V$. 
Upper bounds on the energy scale of inflation impose $V\lesssim 10^{16}~\mathrm{GeV}$~\cite{Hertzberg:2008wr,Aghanim:2018eyx}. 
A short time after its formation, the string system enters into a ``scaling" regime, in which the population of strings in a Hubble volume tends to remain of $\mathcal{O}(1)$ (see e.g. Ref.~\cite{Vilenkin:1984ib} and references therein).

The second term in Eq.~\eqref{eq:potential} explicitly breaks the $U(1)$ symmetry of the first term into a $Z_{N}$ discrete subgroup, producing $N$ degenerate vacua along the previous $U(1)$ orbit of minima $|\phi|\simeq V$. Close to each minimum the ALP field $a$ has a mass
\begin{equation} \label{m-a}
m_a \simeq \frac{ v^2 N}{\sqrt{2} V}.
\end{equation}
We assume for simplicity that temperature corrections to $m_a$ are negligible.  We do not expect that these corrections will be important in our scenario since, as we explain below, catastrogenesis is independent of the temperature dependence as it happens when the ALP mass has reached its asymptotic present value.

The equation of motion of the field $a$ in the expanding Universe is that of a harmonic oscillator with damping term $3H \dot{a}$, where $H= (2 t)^{-1}$ is the Hubble expansion rate during the radiation dominated epoch. 
While $3H \gg m_a$ the field does not change with time. As $H$ decreases with time, at a temperature $T_{\rm w}$ when  $H(T_{\rm w}) \simeq m_a/3$,
regions of the Universe with different values of $\theta$ evolve to different minima separated by domain walls of  mass per unit surface (or surface tension)
\begin{equation} \label{sigma}
\sigma \simeq f_\sigma v^2 \frac{V}{N}.
\end{equation}
Here $f_\sigma$ is a model dependent dimensionless parameter. For $N=2$ the model is solvable analytically and $f_\sigma \simeq 6$. Whenever choosing specific values of $N$ and $f_\sigma$ will be required, we will assume $N=6$ and $f_\sigma/N \simeq 1$.  

 The temperature $T_{\rm w}$ at which walls appear, i.e.
 $H(T_{\rm w}) \simeq m_a/3$, is
\begin{equation} \label{Tw}
T_{\rm w} \simeq \frac{1.6 \times 10^9 \, {\rm GeV}}{\left[g_\star(T_{\rm w})\right]^{1/4}}  \left(\frac{m_a}{\rm GeV}\right)^{1/2} ,
\end{equation}
where we have used that for a radiation dominated Universe the  Hubble expansion rate is 
\begin{equation} \label{def-T}
H(T) = \sqrt{\frac{8 \pi^3 g_\star(T)}{90}}~ \frac{T^2}{M_{\rm P}}.
\end{equation}
Here,  $M_{\rm P}= 1.22 \times 10^{19}$ GeV is the Planck mass and  $g_\star$ is the energy density number of degrees of freedom~\cite{Saikawa:2018rcs}. 

A short time after walls form,  when friction of the walls with the surrounding medium is negligible,  the string-wall system enters into another scaling regime in which the linear size of the walls is the cosmic horizon size $\simeq t$. Therefore, its energy density at time $t$ is
\begin{equation} \label{rhow}
\rho_{\rm wall} \simeq \frac{\sigma}{t}.
\end{equation}
Friction on walls, due to scattering with particles in the surrounding medium, is negligible when the energy of these particles is much larger than the ALP mass $m_a$~\cite{Huang:1985tt, Blasi:2022ayo}, which is the case for most if not all the models we consider. This has been confirmed e.g. in Ref.~\cite{ZambujalFerreira:2021cte}. There the impact of friction on the parameter space of the high-quality QCD axion was studied (see their appendix D), and it was concluded that even when friction was important their results were only mildly different from those obtained neglecting friction.

The third term in Eq.~\eqref{eq:potential}, proposed in Ref.~\cite{Sikivie:1982qv}, is assumed to be much smaller than the second one, i.e. $\epsilon_b \ll~1$. 
It breaks explicitly the $Z_{N}$ symmetry, raising all vacua with respect to the lowest energy one by a bias of order   
\begin{equation} \label{Vbias}
V_{\rm bias} \simeq~\epsilon_b v^4~.
\end{equation}

 The motion of the walls after they are formed is determined by two quantities. The surface tension $\sigma$ tends to straighten out curved walls to the horizon scale as it produces a pressure $p_T\simeq \rho_{\rm wall}$, while the bias  produces a volume pressure $p_V \simeq V_{\rm bias}$~\cite{Gelmini:1988sf}.  This latter pressure  tends to accelerate the walls towards their higher energy adjacent vacuum, converting the higher energy vacuum into the lower energy one, which releases energy that fuels the wall motion. 
 
 Assuming  that  $p_V \ll  p_T$ when walls form (i.e. $\epsilon_b \ll 1$), as  $p_T$ decreases with time it becomes similar to and then smaller than $p_V$. At this point the bias drives the walls (and strings bounding them) to annihilation.  
 Taking $p_T \simeq p_V$, i.e. $\rho_{\rm wall} \simeq \sigma/t_{\rm ann} \simeq V_{\rm bias} \simeq \epsilon_b v^4$, as the condition for wall annihilation, 
\begin{equation} \label{def-Tann}
H(T_{\rm ann})= \frac{1}{2 t_{\rm ann}} \simeq \frac{V_{\rm bias}}{2 \sigma} \simeq \frac{\epsilon_b m_a}{\sqrt{2} f_\sigma},
\end{equation}
which defines the temperature $T_{\rm ann}$  at which the string-wall system annihilates, 
\begin{equation} \label{Tann}
 T_{\rm ann} \simeq \frac{1.9 \times 10^9 ~{\rm GeV}}{[g_\star (T_{\rm ann})]^{1/4}}~  \sqrt{\frac{V_{\rm bias}}{\sigma ~ {\rm GeV}}} \simeq \frac{2.2 \times 10^9  ~{\rm GeV}}{[g_\star(T_{\rm ann})]^{1/4}}~  \sqrt{\frac{\epsilon_b ~ m_a}{f_\sigma ~ {\rm GeV}}}.
\end{equation}
At this point the energy stored in the string-wall system goes almost entirely into non-relativistic or mildly relativistic ALPs (since the wall thickness is $\simeq m_a^{-1}$)~\cite{Chang:1998tb}, into GWs, and into PBHs.

The GWs produced at annihilation (see Appendix~\ref{sec:appendix} and e.g. Ref.~\cite{Gelmini:2022nim} for more details) have a characteristic spectrum peaked at 
\begin{equation} \label{f-peak}
f_{\rm peak}\simeq  0.76~ \text{Hz} ~\frac{T_{\rm ann}}{10^7~{\rm GeV}}~ \frac{\left[g_\star(T_{\rm ann})\right]^{1/2}}{\left[g_{s \star}(T_{\rm ann}) \right]^{1/3}},
\end{equation}
and peak energy density (see e.g.~~\cite{Hiramatsu:2010yz,Hiramatsu:2013qaa,Kawasaki:2011vv,Hiramatsu:2012sc})
\begin{equation} \label{eq:OmegaGW-walls}
\Omega_{\rm GW}h^2|_{\rm peak} \simeq \epsilon_{\rm GW}
\frac{1.2 \times 10^{-79} g_\star(T_{\rm ann})~ \sigma^4}{\left[g_{s \star}(T_{\rm ann}) \right]^{4/3} V_{\rm bias}^2{\text{GeV}}^4}
\simeq \epsilon_{\rm GW}
\frac{1.2 \times 10^{-79}  g_\star(T_{\rm ann})}{\epsilon_b^{2}~\left[g_{s \star}(T_{\rm ann}) \right]^{4/3}}
\left(\frac{f_\sigma V}{N {\text{GeV}}}\right)^4 .
\end{equation}
Following Ref.~\cite{Hiramatsu:2012sc},  we include  in Eq.~\eqref{eq:OmegaGW-walls} a dimensionless factor $\epsilon_{\rm GW} \simeq$  10 -- 20 for $N=6$ (see Fig.~8 of Ref.~\cite{Hiramatsu:2012sc}) found in numerical simulations that parameterizes the efficiency of GW production. In the following we conservatively assume $\epsilon_{\rm GW}=10$.  

Eq.~\eqref{eq:OmegaGW-walls} is also (see Appendix~\ref{sec:appendix}) the usually quoted  maximum of the GW energy density spectrum at present as a function of the wave-number at present $k$, which for the present scale factor $R_0=1$ coincides with the comoving wave-number, or of the frequency $f=k/(2\pi)$. This is defined as
$\Omega_{\rm GW}h^2 (k, t_0) = \left[h^2/{\rho_c(t_0)}\right]
 \left({d \rho_{\rm GW} (t)}/{d \ln k}\right)_0= \left[{h^2}/{\rho_c(t_0)}\right]  \sigma^2/M_{\rm P}^2$~\cite{Maggiore:1900zz,Gelmini:2020bqg}.

 GWs are also emitted by the string network before walls appear. The dominant source of these waves are loops continuously formed by string fragmentation. 
  From the spectra computed in Refs.~\cite{Chang:2019mza,Gouttenoire:2019kij,Gorghetto:2021fsn},  also based on the quadrupole formula, a simple good fit  can be derived for the energy spectrum of GWs emitted
 by strings during the radiation dominated era, namely
\begin{equation} 
\label{OmegaGW-strings}
\Omega^{\rm st}_{\rm GW} h^2
\simeq 3.6 \times 10^{-9} \left(\frac{10^{2} ~ {\rm Hz}}{f} \right)^{1/8} \left( \frac{V}{10^{16}~ {\rm GeV}} \right)^4.
\end{equation}
This spectrum has a low frequency cutoff $f^{\rm st}_{\rm cut}$ at the frequency of GWs emitted the latest possible by strings, when the horizon size is largest, namely when walls are formed, which is given by Eq.~\eqref{f-peak} using $T_{\rm w}$ instead of $T_{\rm ann}$, i.e.
\begin{equation}
f^{\rm st}_{\rm cut} 
 \simeq \frac{T_{\rm w}}{T_{\rm ann}} f_{\rm peak}
\frac{\left[g_\star(T_{\rm w})\right]^{1/2}}{\left[g_{s \star}(T_{\rm w}) \right]^{1/3}}  \frac{\left[g_{s \star}(T_{\rm ann}) \right]^{1/3}}{\left[g_\star(T_{\rm ann})\right]^{1/2}},
\end{equation}
which can be written in terms of the ALP mass as
\begin{equation}\label{fst-cut}
f^{\rm st}_{\rm cut} 
 \simeq 82~ {\rm Hz} \left(\frac{m_a}{\rm GeV}\right)^{1/2}  
\left[\frac{g_\star(T_{\rm w})}{105}\right]^{1/4} 
\left[\frac{105}{g_{s\star}(T_{\rm w})}\right]^{1/3}.
\end{equation}
Given that we consider only ALPs with mass $m_a\gtrsim 1\,\rm GeV$, the emission of GWs by strings before walls appear only contributes at high frequencies, $f \gtrsim 82\,\rm Hz$. Comparing Eqs.~\eqref{OmegaGW-strings} and~\eqref{eq:OmegaGW-walls} we are going to show that only in a restricted
parameter space GWs from strings could be observable by the future GW Einstein Telescope~\cite{Sathyaprakash:2012jk}.

In the next section we will argue that when ALPs are unstable and decay fast in the early Universe, the string-wall system annihilation could produce PBHs that account for the whole (or part) of the dark matter, while also producing GWs with peak frequency from $10^{-5}$ to $10^{2}\,\rm Hz$ and density $\Omega_{\rm GW}h^2 > 10^{-15}$, which corresponds to an observable signal in future GW detectors.

\section{Primordial black holes}
\label{PBH}

In $N=1$ models the walls appear in the form of ribbons flanked by two strings which become narrower due to surface tension and disappear very fast. By contrast, in the latest stages of wall annihilation  in $N>1$ models, closed walls are expected to arise and collapse in an approximately spherically symmetric way. In this case, some fraction of the closed walls could shrink to their Schwarzschild radius $R_{\rm Sch}(t)= 2GM(t)= 2 M(t)/M_{\rm P}^2$ and collapse into PBHs~\cite{Ferrer:2019pbh}.  Here $M(t)$ is the mass within the collapsing closed wall at time $t$. Considering that during the scaling regime the typical linear size of the walls is the horizon size, which we take to be $\simeq t$, 
 PBH formation could happen if the ratio
\begin{equation}
\label{eq:ptRadius}
    p(t)= \frac{R_{\rm Sch}(t)}{t}  =\frac{2 M(t)}{t\, M_{\rm P}^2}
\end{equation}
is close to one, i.e. $p(t)\simeq 1$. 
Simulations of the annihilation process~\cite{Kawasaki:2014sqa} can then be used to estimate at which temperature
$T_\star\lesssim T_{\rm ann}$ this could happen.

As we mentioned above, the annihilation process starts at $T_{\rm ann}$ when the contribution of the volume energy density to the mass within a closed wall of radius $t$  becomes as important as the contribution of the wall energy density. Shortly after, the volume density term dominates over the surface term, and the volume pressure accelerates the walls towards each other. 
 Close to the start of the annihilation process,  the mass within a closed wall as a function of the lifetime $t$ of the Universe is 
\begin{equation} 
    M(t)\simeq \frac{4}{3}\pi t^3 V_{\rm bias} + 4\pi t^2 \sigma.
    \label{Mt}
\end{equation}
Therefore, the ratio $p(t)$ increases with time (as $t^2$ once the volume term becomes dominant). At $T >  T_{\rm ann}$ the surface term dominates and at $T < T_{\rm ann}$ the volume term dominates. If at the moment $t_{\rm ann}$ when annihilation starts $p(t_{\rm ann})$ is  close to 1, PBHs would form immediately. If $p(t_{\rm ann}) \ll 1$ instead, PBHs could only form later, at a  time $t_\star > t_{\rm ann}$ corresponding to a temperature $T_\star < T_{\rm ann}$, for which $p(t_\star)=1$. Temperature and time are related by $H=1/2t$ (see Eq.~\eqref{def-T}),  assuming radiation domination. If $p(t_{\rm ann}) \ll 1$, at  $T_\star$ only a small fraction of the original string-wall system still remains. 

 The  wall surface energy density  $\sigma/t$ decreases with time with respect to the volume energy density due to the bias $V_{\rm bias}$ which is constant in time. Although the volume energy density is negligible initially, both contributions become similar at $T_{\rm ann}$, i.e. $V_{\rm bias} \simeq\sigma/t_{\rm ann}$, thus Eq.~\eqref{Mt} implies  
\begin{equation}
M(t_{\rm ann})\simeq \frac{16}{3} \pi t_{\rm ann}^3 V_{\rm bias},
\end{equation}
 so that the ratio $p(t)$ in Eq.~\eqref{eq:ptRadius} becomes
\begin{equation}
   p(T_{\rm ann})\simeq 
   \frac{30}{\pi^2} \frac{V_{\rm bias}}{g_\star(T_{\rm ann})~T_{\rm ann}^4}.
\end{equation}
As expected of all quantities depending only on the annihilation process, $M(t_{\rm ann})$ and $p(T_{\rm ann})$ only depend on the parameters $V_{\rm bias}$ and $\sigma$.  After $t_{\rm ann}$,  the volume contribution to the density rapidly becomes dominant over the surface contribution, and 
\begin{equation} \label{eq:M-late}
    M(t)\simeq \frac{4}{3}\pi t^3 V_{\rm bias} \left(1 + 3 \frac{t_{\rm ann}}{t}\right),
\end{equation}
thus
\begin{equation} \label{eq:p-late}
 p(T)\simeq \frac{p(T_{\rm ann})}{4} 
 \left(\frac{t}{t_{\rm ann}}\right)^2 \left( 1 + 3 \frac{t_{\rm ann}}{t} \right).
\end{equation}
When $t\gg 3 ~t_{\rm ann}$ we can
 we neglect the second term in Eqs.~\eqref{eq:M-late} 
 and~\eqref{eq:p-late}, consequently 
\begin{equation}
 p(T)\simeq \frac{p(T_{\rm ann})}{4} \frac{g_\star(T_{\rm ann})}{g_\star(T)}  \left(\frac{T_{\rm ann}}{T}\right)^4.
 \label{pT<Tann}
\end{equation}
We are here assuming that the characteristic linear dimension of the walls continues to be close to $t$ even after annihilation is underway. At some point along this process one expects larger deviations from this scaling behavior, but detailed simulations that are not available at the moment would be needed to assess when and how this departure happens. Using Eq.~\eqref{pT<Tann} we find the PBH formation temperature $T_\star$ as the temperature at which $p(T_\star)=1$, which in terms of $T_{\rm ann}$ is
\begin{equation}
 p(T_\star) \simeq \frac{p(T_{\rm ann})}{4}  \frac{g_\star(T_{\rm ann})}{g_\star(T_\star)} \left(\frac{T_{\rm ann}}{T_\star}\right)^4 =1 .
 \label{Eq:p(T)}
\end{equation}
This relation defines $T_\star$ and its corresponding time $t_\star = 1/2 H(T_\star)$. The PBH mass is then $M(t_\star)$, 
\begin{equation} \label{eq:MPBH}
    M_{\rm PBH} = M(t_\star) \simeq \frac{4 \pi}{3}V_{\rm bias} t_\star^3 \simeq \frac{2}{[p(T_{\rm ann})]^{3/2}} M(t_{\rm ann}) \simeq \left(\frac{3}{32 \pi}\right)^{1/2} \frac{M_{\rm P}^3}{V_{\rm bias}^{1/2}} .
\end{equation}
Eqs.~\eqref{Eq:p(T)} and~\eqref{eq:MPBH} show that the temperature of PBH formation $T_\star$ depends only on $V_{\rm bias}$ (or, equivalently,  $ M_{\rm PBH}$)
\begin{equation}
T_\star \simeq 0.9~ {\rm GeV} \left[\frac{V_{\rm bias}}{{\rm GeV}^4~g_\star(T_\star)}\right]^{1/4} \simeq \frac{0.5 ~ {\rm GeV}}{[g_\star(T_\star)]^{1/4}} \left(\frac{M_\odot}{M_{\rm PBH}}\right)^{1/2}.
\label{T star}
\end{equation}
From Eq.~\eqref{eq:MPBH} and Eq.~\eqref{eq:ptRadius} one finds
\begin{equation}
  p(T_{\rm ann})\simeq \frac{t_{\rm ann}^2M_{\rm P}^4}{M_{\rm PBH}^2}  =
   \frac{90}{32 \pi^3}\frac{1}{g_\star(T_{\rm ann})}  \frac{M_{\rm P}^6}{T_{\rm ann}^4 M_{\rm PBH}^2} = \frac{2.4 \times 10^{-33}}{g_\star (T_{\rm ann})}
    \left(\frac{\rm GeV}{T_{\rm ann}}\right)^4
    \left(\frac{10^{16} M_\odot}{M_{\rm PBH}}\right)^2.
   \label{eq:pann-1}
\end{equation}

Small deviations from a spherically symmetric collapse, as well as effects of some angular momentum, could make the probability of forming a PBH at temperature $T$ somewhat  smaller than $p(T)$, which we could attempt to parameterize  as $p(T)^\beta$ with a real positive exponent $\beta$. A modification of the probability  of this type does not have any effect on our estimates of PBH formation, since when $p=1$ also $p^\beta =1$. A large enough deviation from the spherical shape could prevent the formation of a PBH, since the degree of asymmetry may decrease initially during the contraction but increase in the late stages of the collapse~\cite{Widrow:1989fe}. The details of the collapse process become more important if $p(T_{\rm ann}) \ll 1$, since a longer evolution is needed to potentially reach $p=1$ (Ref.~\cite{Ferrer:2019pbh} suggests that in this case  asphericities, energy losses or angular momentum might make more difficult the formation of a black hole). However, a large deviation from sphericity is unlikely, as shown in the context of the collapse of vacuum bubbles produced during inflation~\cite{Deng:2017uwc}. Thus, at least for some portion of the walls, the departure from spherical symmetry during collapse is likely to be small. 
In this case, the PBH density at formation is given by the energy density in the wall system when PBHs form times the probability of PBH formation, namely
\begin{equation} 
\rho_{\rm PBH}(T_\star) \simeq  p^\beta(T_\star) \rho_{\rm wall}(T_\star),
\end{equation}
and the  fraction $f_{\rm PBH}$ of the total  DM density $\rho_{\rm DM}$ in PBHs is 
\begin{equation} 
\label{eq:PBH fraction}
    f_{\rm PBH}=\frac{\rho_{\rm PBH}(T_\star)}{\rho_{\rm DM}(T_\star)}     \simeq p^\beta(T_\star)\frac{\rho_{\rm wall}(T_\star)}{\rho_{\rm DM}(T_\star)}=
    \frac{\rho_{\rm wall}(T_\star)}{\rho_{\rm wall}(T_{\rm ann})}
    \frac{\rho_{\rm wall}(T_{\rm ann})}{\rho_{\rm DM}(T_\star)},
\end{equation}
where we used that at the moment of PBH formation temperature $p(T_\star)=1$.
Notice that $f_{\rm PBH}$ is the same at present as it was at $T_\star$ (since the PBH and DM densities redshift in the same manner).

Simulations of the process of the string-wall system annihilation for $N>1$ models have been performed~\cite{Kawasaki:2014sqa}, which provided measurements of the
times at which the area density of walls (area per unit volume $A/V$) is 10\% and 1\% of what it would have been without a bias. Notice that this ratio is the same in comoving (as given in Ref.~\cite{Kawasaki:2014sqa}) or physical coordinates. We call these times $t(10\%)$ and $t(1\%)$, and the corresponding temperatures $T(10\%)$ and $T(1\%)$. In Ref.~\cite{Kawasaki:2014sqa}, the pressure due to the walls is parameterized as $\mathcal{A}(t) \sigma/t$, where $\mathcal{A}(t)$ is a function which accounts for deviations from scaling, and find that $\mathcal{A}(t)$ is close to 1.  Assuming that the wall contribution to the energy density is dominant, a simple approximation as a power law for the evolution of the wall energy density with temperature after annihilation starts,  
\begin{equation}
   \frac{\rho_{\rm wall}(T)}{\rho_{\rm wall}(T_{\rm ann})}=
    \left(\frac{T}{T_{\rm ann}} \right)^\alpha ,
    \label{Eq Def of alpha}
\end{equation}
seems to work well to extract the real positive exponent $\alpha$ from the mentioned simulations~\cite{Kawasaki:2014sqa}.  

Table~VI  and Fig.~4 of Ref.~\cite{Kawasaki:2014sqa} show that the $t(1\%)/t(10\%)=[T(10\%)/T(1\%)]^2$  ratio takes up central values close to 2, actually from 1.7 to 1.5, under different assumptions. If the energy of the string-wall system is still dominated  by the contribution of the walls until $t(1\%)$, i.e. the energy density is proportional to $A/V$, $\rho_{\rm wall}= A \sigma/V$, then  
\begin{equation}
   \left(\frac{T(10\%)}{T(1\%)} \right)^\alpha \simeq \frac{\rho_{\rm wall}(T(10\%))}{\rho_{\rm wall}(T(1\%))}
   = 
 \frac{\left. \frac{A}{V}\right|_{t(10\%)}}{ \left.\frac{A}{V}\right|_{t(1\%)} } = 10 ~ \left(\frac{t(1\%)}{t(10\%)}\right),
    \label{Eq Def of alpha-2}
\end{equation}
  and the central values measured of the $t(1\%)/t(10\%)$  ratio translate into values of the exponent $\alpha$ roughly between 9 and 14. In our figures we include also the value 7 used in Ref.~\cite{Ferrer:2019pbh}.
  Taking into account the systematic errors quoted in Table~VI  of  Ref.~\cite{Kawasaki:2014sqa}, $\alpha$ could be as large as 21.
  However, the volume contribution to the energy density of the string-wall system may not be negligible, which introduces a further uncertainty in the determination of $\alpha$. To get an estimate of this uncertainty, we can proceed assuming that the volume energy is dominant in the string-wall system, i.e. that its density is proportional to $A^{3/2}$. In fact, since the simulation volume in Ref.~\cite{Kawasaki:2014sqa} is the same in both the cases with and without a bias (in the latter case the evolution follows the scaling solution), the ratio of the area densities can be identified with the ratio of the area $A$ in the evolution with bias and the characteristic area $\simeq t^2$  in the scaling case within a Hubble volume. Thus $[A(t(10\%))]^{1/2} \simeq \sqrt{0.10}~ t(10\%)$,  and similarly $[A(t(1\%))]^{1/2}\simeq \sqrt{0.01} ~ t(1\%)$. Moreover, in a Hubble volume, if  the $V_{\rm bias}$ contribution dominates over the surface energy, the wall-system density is $\rho_{\rm wall} \simeq V_{\rm bias} [A(t)]^{3/2}/ t^3$, and
 \begin{equation}
    \left(\frac{T(10\%)}{T(1\%)} \right)^\alpha \simeq \frac{\rho_{\rm wall}(T(10\%))}{\rho_{\rm wall}(T(1\%))}\simeq 10^{3/2} .
    \label{Eq Def of alpha-3}
\end{equation}
 This leads to values of $\alpha$ larger by a factor $3/2$ with respect to our previous estimates minus two (i.e. from $(3/2)7$ to $(3/2)19$). Thus, $\alpha$ could be as large as 28.  
 
Using Eq.~\eqref{Eq Def of alpha}, Eq.~\eqref{eq:PBH fraction}  becomes
\begin{equation} 
\label{eq:PBH fraction-2}
    f_{\rm PBH} \simeq
    \left( \frac{T_\star}{T_{\rm ann}}\right)^{\alpha}
    \frac{\rho_{\rm wall}(T_{\rm ann})}{\rho_{\rm DM}(T_\star)}.
\end{equation}
In this equation $\rho_{\rm DM}(T_\star)$ is easily related by redshift to the present dark matter density, and $\rho_{\rm wall}(T_{\rm ann})$ can be related to the present radiation density, in the following manner.

Had the string-wall system not annihilated, its energy density would have continued evolving in the scaling regime, with $\rho_{\rm wall} \simeq \sigma/ t$, until the time $t_{\rm wd}$ at which $\rho_{\rm wall}(t_{\rm wd})= \rho_{\rm rad}(t_{\rm wd})$,  thus 
\begin{equation}
\rho_{\rm wall}(t_{\rm ann}) = \frac{t_{\rm wd}}{t_{\rm ann}}~ \rho_{\rm rad}(t_{\rm wd})
= \frac{H(T_{\rm ann})}{H(T_{\rm wd})}~\rho_{\rm rad}(t_{\rm wd}).
\end{equation}
Assuming radiation domination up to this point, the wall-domination temperature $T_{\rm wd}$ is
\begin{equation}
\label{T wall domination}
    T_{\rm wd}\simeq \frac{0.9 \times 10^{-9}~{\rm GeV}}{\left[g_\star(T_{\rm wd})\right]^{1/4}} \left(\frac{\sigma}{{\rm GeV}^3}\right)^{1/2} \simeq
    \frac{2.0~{\rm GeV}~[g_\star(T_\star)]^{1/2}}{[g_\star(T_{\rm wd}) g_\star(T_{\rm ann})]^{1/4}}
    \left(\frac{T_\star}{\rm GeV}\right)^2 \left(\frac{\rm GeV}{T_{\rm ann}}\right),
\end{equation}
or
\begin{equation}
\label{T wall domination-2}
    T_{\rm wd} \simeq 
    \frac{3.4~ {\rm GeV}}{[g_\star(T_{\rm wd})]^{1/4}}
    \frac{f_\sigma^{1/2}}{N}\left(\frac{V}{\rm 10^9~GeV}\right)
  \left(\frac{m_a}{\rm 10~GeV}\right)^{1/2}.
\end{equation}
Considering the redshift of the radiation density to the present, Eq.~\eqref{eq:PBH fraction-2} becomes
\begin{equation}
    f_{\rm PBH} \simeq
    \frac{g_{s\star}(T_0)}{g_\star(T_0)} ~
    \frac{\left[g_\star(T_{\rm ann}) g_\star(T_{\rm wd})\right]^{1/2}}{g_{s\star}(T_\star)} ~ \frac{T_\star^{(\alpha -3)} ~ T_{\rm wd}^2}{T_0 ~ T_{\rm ann}^{(\alpha-2)}} \left(\frac{\rho_{\rm rad}}{\rho_{\rm DM}}\right)_0 ,
\end{equation}
or, using Eq.~\eqref{T wall domination}, 
\begin{equation}
    f_{\rm PBH} \simeq
    \frac{4.0~g_{s\star}(T_0)}{g_\star(T_0)} ~
    \frac{g_\star(T_\star)}{g_{s\star}(T_\star)} ~ \frac{T_\star^{(\alpha +1)}}{T_0 ~ T_{\rm ann}^\alpha }~ \left(\frac{\rho_{\rm rad}}{\rho_{\rm DM}}\right)_0 .
\end{equation}
Using Eq.~\eqref{T star} and the values of the present quantities in the previous equation we find
\begin{align}
  \nonumber  f_{\rm PBH} \simeq & 1.1 \times 10^{14} \times 1.5^\alpha \times  10^{-2\alpha} \left(\frac{10^{-10} \, M_{\odot}}{M_{\rm PBH}} \right)^{(\alpha + 1 )/2}  \left( \frac{10^6 \, \mathrm{GeV}}{T_{\rm ann}} \right)^\alpha \\
    & \times \left( \frac{105}{g_\star (T_\star)} \right)^{(\alpha - 3)/4}\left( \frac{105}{g_{s\star} (T_\star)} \right)
    \label{fPBH vs Mpbh Tann}.
\end{align}
In the following we neglect the possible change of energy degrees of freedom between $T_{\rm ann}$ and $T_\star$, since these temperatures are always close to each other.

\begin{figure}  
  \centering 
  \includegraphics[width=0.85\linewidth]{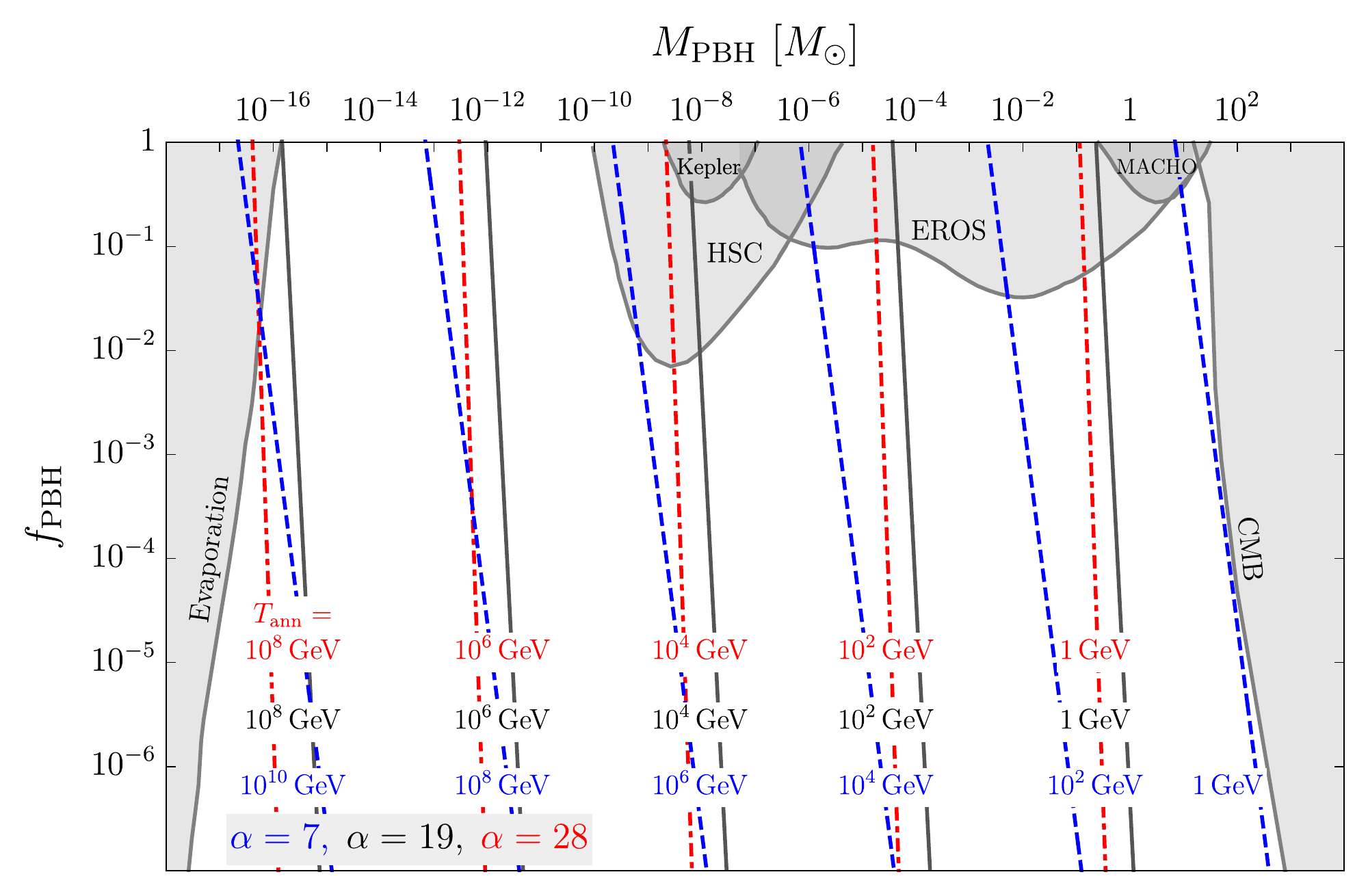}
\caption{Fraction of the DM in PBH as a function of the PBH mass (assuming a monochromatic mass function) for different values of the string-wall system annihilation temperature ${T_{\rm ann}}$ (see Eq.~\eqref{fPBH vs Mpbh Tann}) and of the parameter $\alpha$  (7 for dashed blue lines, 19 for solid gray lines and 28 for dash dotted red lines). Observational upper limits  on $f_{\rm PBH}$ are shown in gray (see main text for details). 
}
\label{Fig fPBH MPBH}
\end{figure}

\begin{figure}  
  \centering
  \includegraphics[width=0.85\linewidth]{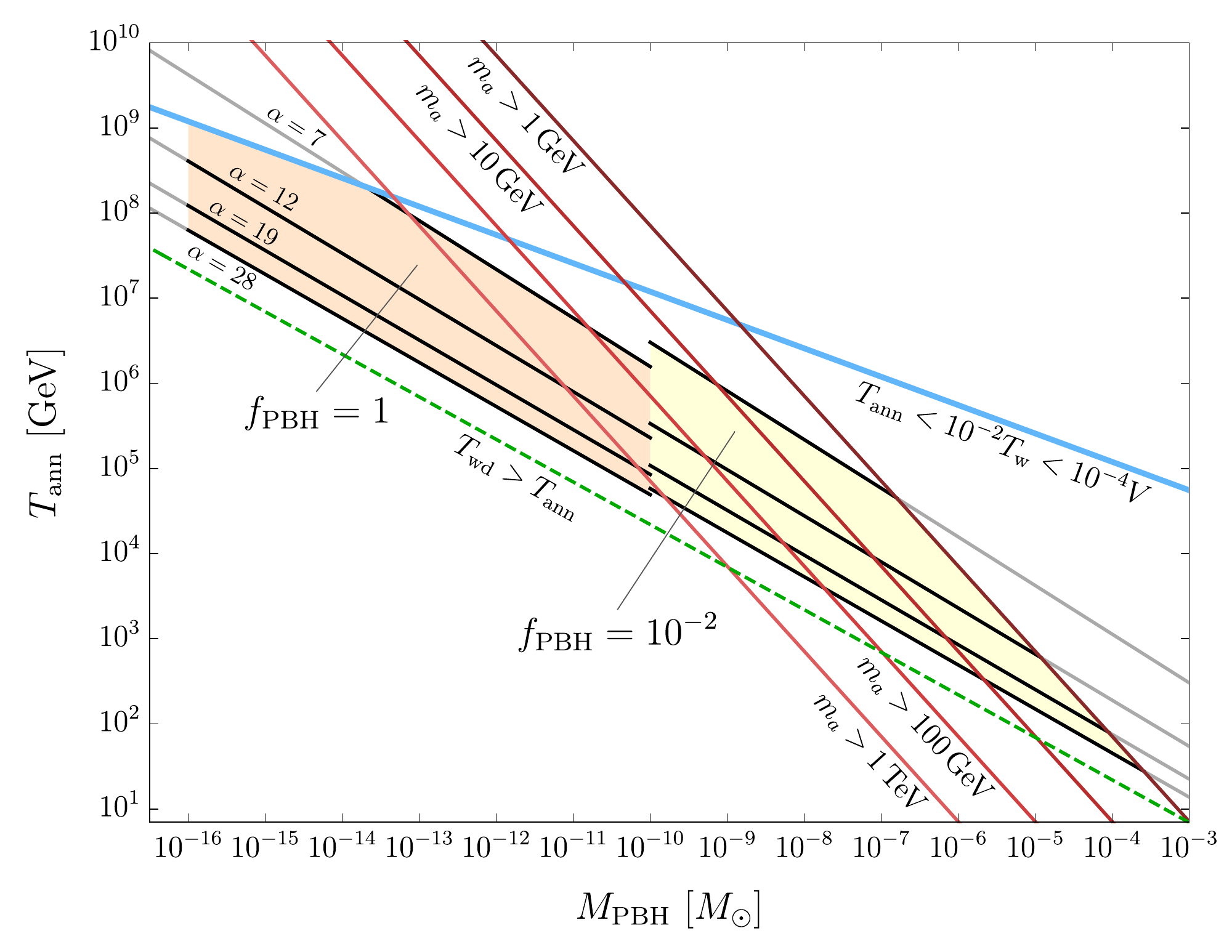}
\caption{Range of ${T_{\rm ann}}$  as function of PBH mass for which
Eq.~\eqref{fPBH vs Mpbh Tann} implies that PBHs can account for all of the DM, i.e. $f_{\rm PBH}=1$ (orange band) or $f_{\rm PBH}=10^{-2}$ (yellow band) which is allowed by all upper limits (as shown in Fig.~\ref{Fig fPBH MPBH}). Each lower limit on $m_a$ excludes the region above and to the right of the corresponding red line, imposing for consistency of the model that $T_{\rm w}< 10^{-2}\,V$ (i.e. $m_a$ can only be larger than the quoted limit to the left of the lines, and can be both smaller on both sides). 
Other consistency conditions,  $T_{\rm ann} <10^{-2}~ T_{\rm w} < 10^{-4}~ V$ and $T_{\rm wd} < T_{\rm ann}$, reject the regions above the thick blue line and below the dashed green line, respectively. 
}
\label{Fig Tann MPBH}
\end{figure}

Fig.~\ref{Fig fPBH MPBH} shows upper limits on the fraction of DM in PBH~$f_{\rm PBH}$ as a function of the PBH mass~$M_{\rm PBH}$. The constraints correspond to Hawking's evaporation bounds~\cite{Carr:2009jm,Boudaud:2018hqb,Laha:2019ssq,DeRocco:2019fjq,Laha:2020ivk,Clark:2016nst,Mittal:2021egv}, microlensing bounds~\cite{Griest:2013aaa,Macho:2000nvd,EROS-2:2006ryy,Niikura:2017zjd,Kusenko:2020pcg}, and CMB bounds~\cite{Serpico:2020ehh}. All bounds on the PBH abundance are taken from Ref.~\cite{PBHbounds}, except for the ones from the Subaru Hyper Suprime-Cam (HSC) data~\cite{Niikura:2017zjd,Kusenko:2020pcg}. Additional constraints, independent of cosmology, relies on dwarf galaxy heating~\cite{Lu:2020bmd,Takhistov:2021upb}. The parameter space will be further probed by upcoming experiments~\cite{Saha:2021pqf,Ray:2021mxu}.
The lines of constant annihilation temperature ${T_{\rm ann}}$  implied by Eq.~\eqref{fPBH vs Mpbh Tann} for different values of the parameter $\alpha$ are shown 
in dashed blue for $\alpha=7$, solid gray for $\alpha=19$ and dash dotted red for $\alpha=28$. 
 We can clearly see in the figure that there are annihilation temperatures, in the mass range $10^{-16} M_{\odot}< M_{\rm PBH}  < 10^{-10} M_{\odot}$, for which all of the DM could be in PBHs .

The range  of  ${T_{\rm ann}}$  for which PBHs can constitute 100\% of the DM is better seen as the orange band in Fig.~\ref{Fig Tann MPBH}. The yellow band in the same figure shows the range of  ${T_{\rm ann}}$  in which, according to Eq.~\eqref{fPBH vs Mpbh Tann}
$f_{\rm PBH}=10^{-2}$, a DM fraction that is just below
all the observational bounds for $10^{-10} M_{\odot}< M_{\rm PBH}  < 1 \, M_{\odot}$. Such PBHs could potentially account for some putative events, as found by
HSC~\cite{Niikura:2017zjd} and other~\cite{EROS-2:2006ryy} microlensing observation, as well as LIGO observations~\cite{Sasaki:2016jop}. 

As we explain in Sec.~\ref{sec:consistency}, the consistency requirement of having walls formed well after strings appear, i.e. $T_{\rm w} \ll V$,  implies that a lower limit on $m_a$ translates into a region of this plot being excluded. Assuming $T_{\rm w} < 10^{-2} V$, we obtain the limits shown in Fig.~\ref{Fig Tann MPBH} as red lines.  Each lower limit  on the ALP mass (from 1 GeV to 1 TeV) excludes the region above and to the right of the corresponding red line (meaning that larger values of $m_a$ are only possible to the left of the corresponding line, although also smaller values are possible). 
These limits exclude values of $M_{\rm PBH}$ lower than $10^{-2} M_{\odot}$ for $f_{\rm PBH}=10^{-2}$.  Two combined consistency conditions,  $T_{\rm ann} <10^{-2}~T_{\rm w}$ and $T_{\rm w} < 10^{-2}~ V$, reject the regions above the thick blue line in Fig.~\ref{Fig Tann MPBH}.
Finally, requiring the string-wall system not to dominate the energy density of the Universe, i.e. $T_{\rm wd} < T_{\rm ann}$, rejects the region below the dashed green line, thus it does not affect any of our regions of interest.

Notice that the some of the above mentioned consistency conditions depend on $T_{\rm w}$, and thus on $m_a$. Therefore, they might be affected by a temperature dependence of $m_a$ which we have neglected. However, the results depending only on the string-wall system annihilation, i.e. the present density of PBHs and GWs due to catastrogenesis, are independent of this assumption.

\section{Potentially observable GWs}
\label{GW}

\begin{figure}  
  \centering
  \includegraphics[width=0.9\linewidth]{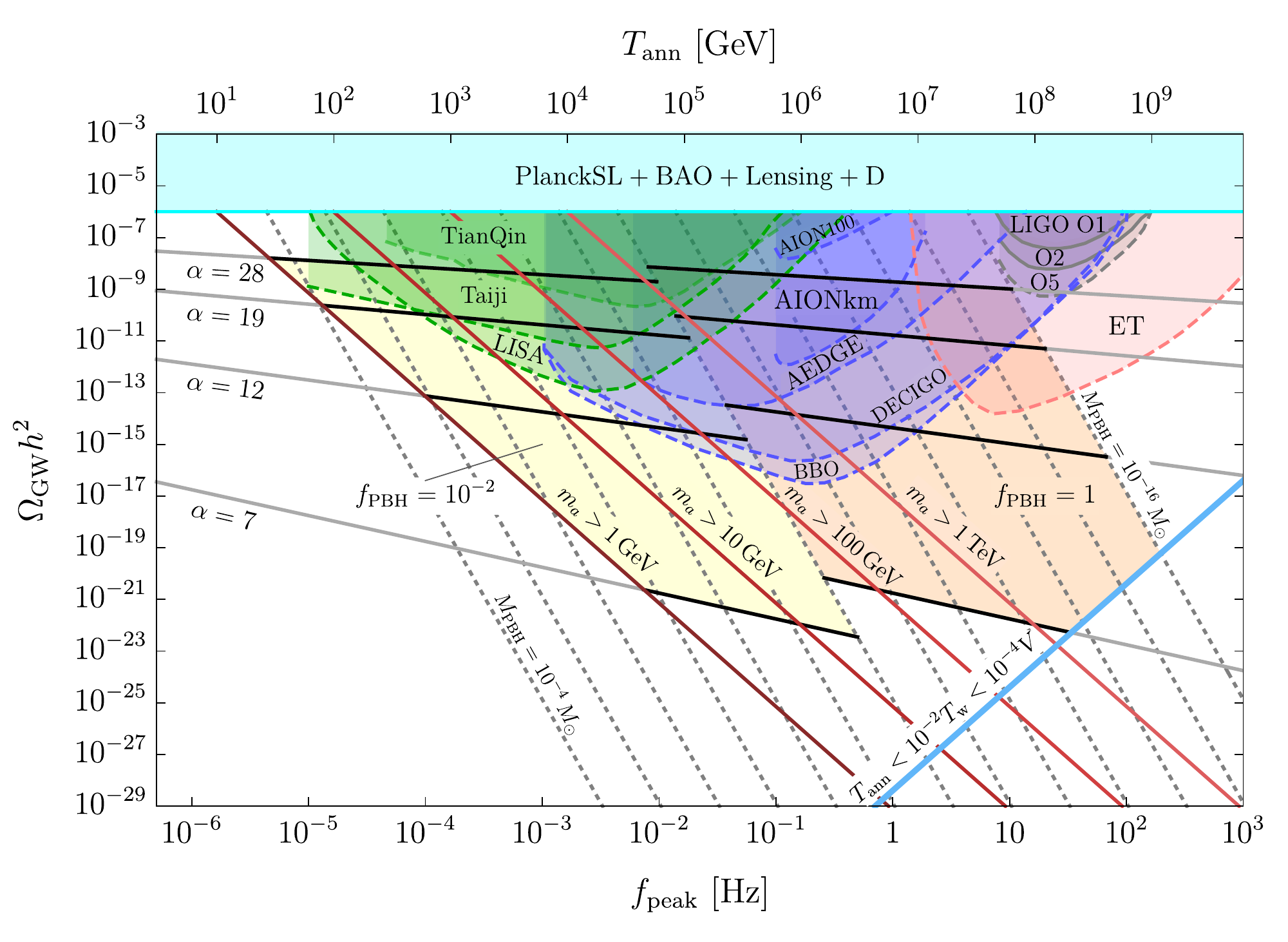}
\caption{Present GW density  $\Omega_{\rm GW} h^2$ emitted by the string-wall system predominantly at annihilation, as a function of  $f_{\rm peak}$ (lower abscissa axis)  or $T_{\rm ann}$ (upper abscissa axis) for $f_{\rm PBH}=1$ (orange region) or  $f_{\rm PBH}=10^{-2}$ (yellow region) expected for different values of the power index $\alpha$, between 7 and 28 (solid black lines). We also show upper limits (solid contours) and the expected reach (dashed contours) of existing and future GW detectors, respectively. 
The quoted lower limits on $m_a$ allow only the regions above and to the right of the respective slanted red lines (meaning that $m_a$ can be larger than the quoted values only to the right side of the lines, and can be smaller on both sides) if $T_{\rm w}/V=10^{-2}$ (for smaller values of this ratio the allowed regions shrink).  Gray dotted lines indicate constant PBH mass values. The consistency condition  $T_{\rm ann} <10^{-2}~ T_{\rm w} < 10^{-4}~ V$  rejects the region below and to the right of the thick blue line.
}
\label{Fig OmegaGW f}
\end{figure}

Since the GWs due to the string-wall system and PBHs are produced at annihilation, their present abundance depends only on two parameters, $V_{\rm bias}$ and $\sigma$, which determine when annihilation happens. It is convenient here to choose the two independent parameters to be instead the GW peak frequency $f_{\rm peak}$ (which depends only on $T_{\rm ann}$ through Eq.~\eqref{f-peak})  and the fraction of DM in PBHs, $f_{\rm PBH}$, and  write $\Omega_{\rm GW}$ as  
\begin{align}
    \nonumber \Omega_{\text{GW}}  & h^2\simeq \left(7.6 \times 10^{-5} \right)^{{\alpha}/({\alpha + 1})} \left(1.4 \times 10^{-7} \right)^{{20}/({\alpha + 1})} \left( {\frac{1 \, \text{Hz}}{f_{\rm peak}}}f_{\rm PBH} \right)^{{8}/({\alpha + 1})}\\
    & \times \left[\frac{105}{g_\star(T_{\rm ann})} 
    \left( \frac{g_{\star}(T_\star)}{105}\right)^2 \right]^{({\alpha - 3})/({\alpha + 1})}
    \left( \frac{105}{g_{s\star}(T_{\rm ann})}\right)^{{4(\alpha + 3)}/{3(\alpha + 1)}}\left(\frac{g_{s\star}(T_\star)}{105}\right)^{{8}/{\alpha + 1}}.
\end{align}

Fig.~\ref{Fig OmegaGW f} shows the expected present GW density produced by the string-wall system as a function of  $f_{\rm peak}$ (lower abscissa axis)  or $T_{\rm ann}$ (upper abscissa axis).
The region where  PBHs can be all of the DM ($f_{\rm PBH}=1$) and the one where they can constitute only a DM subcomponent ($f_{\rm PBH}=10^{-2}$) are shown respectively in orange and yellow. We consider a  range of $\alpha$ values, 7 to 28 (black solid lines). The predictions are compared to the current upper limits (solid contours) and the expected reach (dashed contours) of several GW detectors. We include the
projected sensitivities of the space-based experiments TianQin~\cite{TianQin:2015yph}, Taiji~\cite{Ruan:2018tsw}, and the Laser
Interferometer Space Antenna (LISA)~\cite{LISA:2017pwj} in green, the reach of the Atom Interferometer
Observatory and Network (AION)~\cite{Badurina:2019hst}, the Atomic Experiment for Dark Matter and Gravity Exploration in Space (AEDGE)~\cite{AEDGE:2019nxb}, the Deci-hertz Interferometer Gravitational wave
Observatory (DECIGO)~\cite{Seto:2001qf}, and the Big Bang Observer (BBO)~\cite{Corbin:2005ny} in blue. Finally, we
show in red the reach of the ground based experiments Einstein Telescope (ET) (projection)~\cite{Sathyaprakash:2012jk} and in grey limits and future reach of the Laser Interferometer Gravitational-Wave Observatory (LIGO)~\cite{LIGOScientific:2019vic}. The cyan band corresponds to the 95\% C.L. upper limit on the effective
number of degrees of freedom during CMB emission from Planck, and other data~\cite{Pagano:2015hma}, which imposes $\Omega_{\rm GW} h^2 < 10^{-6}$.

Also shown in Fig.~\ref{Fig OmegaGW f} are the dotted gray lines of constant PBH mass, from 10$^{-16}$ to 10$^{-4}$ $M_\odot$,  stemming from the following relation,
\begin{equation}
    \Omega_{\rm GW} h^2 \simeq 1.4 \times 10^{-9}
     \frac{105~[g_\star(T_{\rm ann})]^3}{\left[g_{s\star}(T_{\rm ann}) \right]^4} \left( \frac{M_\odot}{M_{\rm PBH}} \right)^4 \left( \frac{10^{-7} {\rm Hz}}{f_{\rm peak}} \right)^8.
    \label{eq:OmegaGW-M}
\end{equation}

As explained in the next section, the lower limits quoted  on the ALP mass (between 1 GeV and 1 TeV) allow only the regions above and to the right of the respective slanted red lines (meaning that $m_a$ cannot be larger than the quoted limit to the left)  if $T_{\rm w}/V=10^{-2}$. For smaller values of the
$T_{\rm w}/V$ ratio the allowed regions shrink, i.e. the slanted red lines move to the right.

Fig.~\ref{Fig OmegaGW f} clearly demonstrates that a GW signal due to the annihilation of the cosmic walls is within the reach of several experiments for values of $\alpha$ above about 12, both if PBHs constitute the whole of the DM or just a small fraction of it. 

\begin{figure}  
  \centering 
  \includegraphics[width=0.85\linewidth]{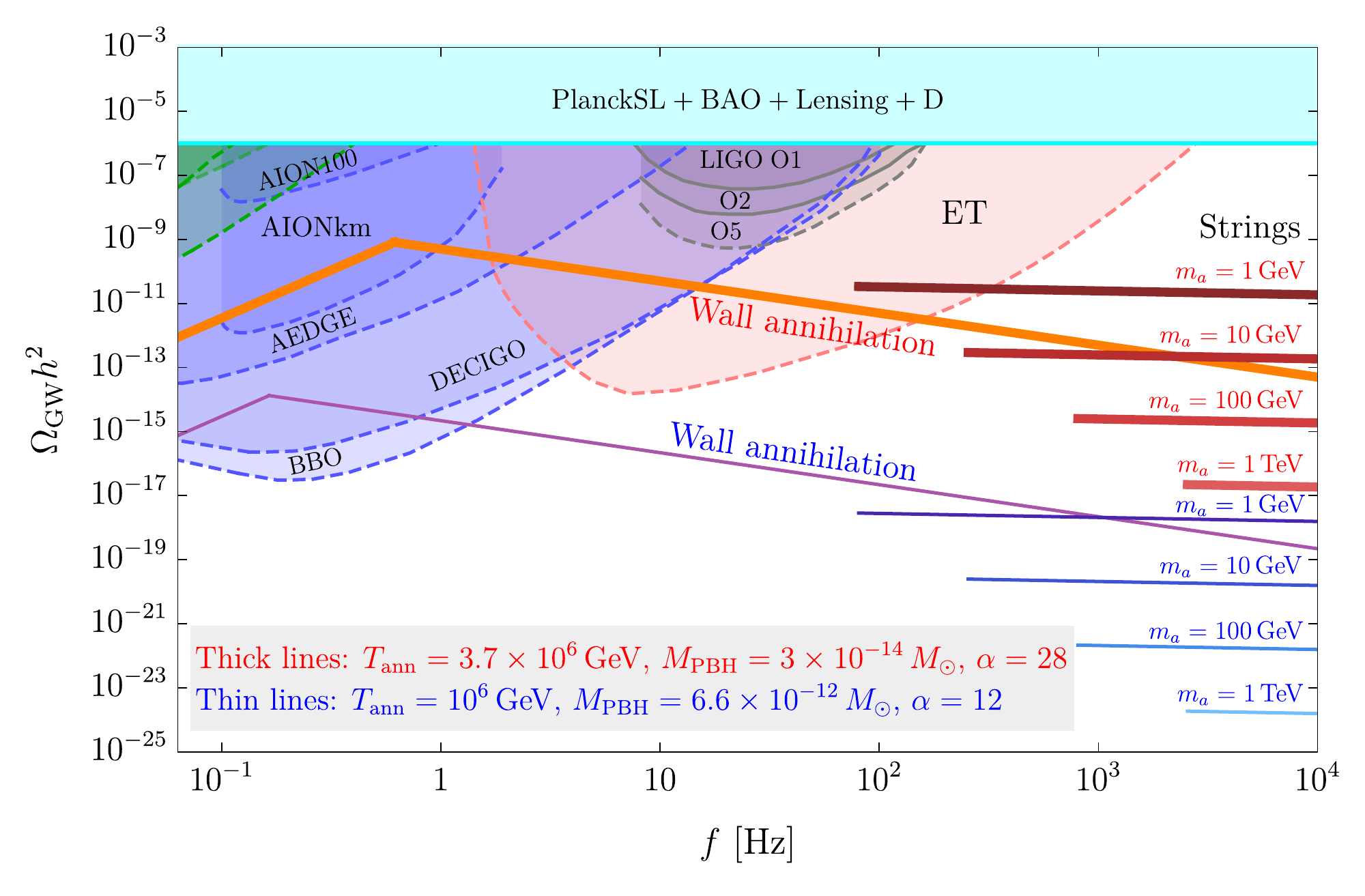}
\caption{Two examples of combined GW spectra from  wall annihilation and from string emission:  for a model with $\alpha= 28$ the spectrum from wall annihilation (thick red solid lines) is observable in several future GW detectors, and the corresponding spectrum from strings (thick dark red lines) for the $m_a$ values indicated  would be observable  by ET  only for $m_a \simeq 1\,\rm GeV$; for a more typical example of a model with lower $\alpha$ value, $\alpha = 12$, the GWs from wall annihilation (in violet) are observable, but the corresponding spectrum from strings (in blue) is not observable. 
}
\label{Fig Omegagwstr}
\end{figure}

Contrary to the GW signal from wall annihilation, the GW signal emitted by strings before walls appear depends on the ALP mass value. Therefore, we do not show the signal from strings in Fig.~\ref{Fig OmegaGW f}. Eq.~\eqref{fst-cut} shows that for $m_a> 1\,\rm GeV$
strings contribute to the GW predicted spectrum only above $f^{\rm st}_{\rm cut}=82$ Hz, and the cutoff frequency moves rapidly to higher frequencies as $m_a^{1/2}$, moving from LIGO frequency range to the range which will be probed by ET. However, Eq.~\eqref{OmegaGW-strings}  shows that the GW energy density due to cosmic strings is always below $10^{-9}$, since we only consider $V< 10^{16}\,\rm GeV$,  thus putting the signal out of the reach of LIGO. The signal is potentially within the reach of ET only for models with very large values of $\alpha$. Fig.~\ref{Fig Omegagwstr} shows two examples of GW spectra from  wall annihilation and from string emission. One of them with $\alpha= 28$ has a large peak amplitude from wall annihilation, with a spectrum $\sim f^3$ below the peak and $\sim 1/f$ above it  (thick red lines), that would be largely observable in several future GW detectors. The corresponding spectrum from strings, shown in dark red for several values of $m_a$, would be observable  by ET  only for $m_a \simeq 1\,\rm GeV$. The other spectra correspond  to more typical examples with lower value of $\alpha$, $\alpha = 12$. In this case the GWs from wall annihilation (shown in violet) could still be observed by DECIGO and BBO, but the GWs from strings (in blue, each for the indicated $m_a$ value) could not. 

Similar predictions for GWs from wall annihilation were found in Ref.~\cite{ZambujalFerreira:2021cte} for the high-quality QCD axion model. Because our consistency conditions are tighter and we include the production of asteroid-mass PBHs, we find a less prominent GW signal. Assuming for example $\alpha=7$, we find that above $1\,\rm Hz$ the GW amplitude cannot be larger than $\Omega_{\rm GW}h^2\simeq 10^{-19}$.

\section{Self consistency conditions} \label{sec:consistency}

Consistency of our models requires that the annihilation of the string-wall system happens before walls would become dominant in the Universe and well after the formation of domain walls, which in turn appear sufficiently after the formation of the cosmic string network, i.e. that $10^{16}~{\rm GeV} > V \gg T_{\rm w} \gg T_{\rm{ann}} > T_{\rm wd}$. Conservatively, we will require $T_{\rm w}/V \lesssim 10^{-2}$ and $T_{\rm{ann}}/T_{\rm w} \lesssim 10^{-2}$ as well.

Our model has three independent parameters, which at the level of the Lagrangian are $V$, $v \ll V$, 
and $\epsilon_b \ll 1$. Instead of $v$, we could alternatively use $m_a$ through Eq.~\eqref{m-a}, with $m_a \ll V$. But for our problem it is convenient to choose as the independent parameters $T_{\rm ann}$ and $M_{\rm PBH}$ (on which anything dependent only on the annihilation depends), as well as $m_a$. In terms of these parameters,
\begin{equation}
        V \simeq \frac{3}{16 \pi} \left( \frac{10}{\pi} \right)^{1/4} \frac{N M_{\rm P}^{7/2}}{f_\sigma^{1/2} \left[g_\star (T_{\rm ann}) \right]^{1/4} m_a^{1/2}T_{\rm ann}M_{\rm PBH}},
\end{equation}
or
\begin{equation}
 V \simeq 1.4 \times 10^{12} \, \mathrm{GeV} \, \frac{N}{f_\sigma^{1/2}} \left(\frac{105}{g_\star (T_{\rm ann})} \right)^{1/4} \left( \frac{1 \, \rm{GeV}}{m_a} \right)^{1/2} \left(\frac{10^6 \, \rm GeV}{T_{\rm ann}} \right) \left( \frac{10^{-10} \, M_{\odot}}{M_{\rm PBH}} 
    \right) \label{eqn:Vhierarchy},
\end{equation}
 and 
 \begin{equation}
    \epsilon_b = 1.2 \times 10^{-5}  \left( \frac{g_\star (T_{\rm{ann}})}{105}\right)^{1/2} \left( \frac{T_{\rm{ann}}}{10^6 \, \mathrm{GeV}} \right)^2 \left( \frac{1 \, \mathrm{GeV}}{m_a} \right) \label{eqn:epsilonb}.
\end{equation}

As shown in Eq.~\eqref{Tw}, $T_{\rm w}$ depends only on $m_a$, since walls appear when $H (T_{\rm w})\simeq m_a/3$. Extracting $m_a$ from the ratio $T_{\rm w}/V$ we can write %
\begin{equation}
 m_a \simeq  29 \, \mathrm{GeV} \, ~\frac{(T_{\rm w}/V)}{10^{-2}}~ \frac{N}{f_\sigma^{1/2}}  \left( \frac{g_\star (T_{\rm w})}{g_\star (T_{\rm ann})}\right)^{1/4} \left( \frac{10^6 \, \rm{GeV}}{T_{\rm ann}}\right) \left( \frac{10^{-10} M_\odot}{M_{\rm PBH}}\right) \label{eqn:Tw/V ma bound}
\end{equation}
from which a lower limit on $m_a$, for a fixed $T_{\rm w}/V$, imposes a limit in the $T_{\rm ann}, M_{\rm PBH}$ plane. The limits so derived are shown as red lines (for $m_a$ larger than 1 GeV, 10 GeV, 100 GeV and 1 TeV) for $T_{\rm w}/V=10^{-2}$ in Fig.~\ref{Fig Tann MPBH}, where they reject the region above and to the right of the lines, and in Fig.~\ref{Fig OmegaGW f}, where they reject the region below and to the left of the lines. As  $T_{\rm w}/V$ decreases, the limits become stronger, rejecting larger regions of the parameter space. 

Notice that using Eq.~\eqref{def-T}, when $T\simeq V$ and strings appear
\begin{equation}
  H(V) = 1.4 \times 10^{-2} \left[\frac{V}{10^{16} \, {\rm GeV}}\right] V \left[\frac{g_\star(V)}{105}\right]^{1/2}~,  
\end{equation}
 and requiring $H (T_{\rm w}) <  H(V)$, i.e. that walls appear sufficiently after strings, implies
\begin{equation}
m_a < 4.2 \times 10^{-2} \left[\frac{V}{10^{16}\,  {\rm GeV}}\right] V 
\left[\frac{g_\star(V)}{105}\right]^{1/2} < 4.2 \times 10^{-2}  V \left[\frac{g_\star(V)}{105}\right]^{1/2}. 
\end{equation}
This shows that the upper limit due to inflation $V< 10^{16}\,\rm GeV$ already implies that $m_a \lesssim 10^{-2} V$. 

Using Eq.~\eqref{def-Tann}, we can easily see that the Hubble parameter at annihilation is $H(T_{\rm ann})= \epsilon_b m_a/ (\sqrt{2} f_\sigma)= \epsilon_b m_a/ 8.46$ (recall that we use $f_\sigma=6$). Thus the condition for annihilation to happen sufficiently after walls appear, i.e. $H(T_{\rm ann})/H(T_{\rm w}) \ll 1$, is equivalent to requiring a small enough $\epsilon_b$ value. Notice that since $H (T_{\rm w})\simeq m_a/3$ and $H(T_{\rm ann})/H(T_{\rm w})= (T_{\rm ann}/ T_{\rm w})^2$, 
  \begin{equation}
  \epsilon_b = 2.8 \left(\frac{ T_{\rm ann}}{T_{\rm w}}\right)^2.   
  \end{equation}
Thus having $\epsilon_b$ small enough is the same as having $T_{\rm ann} \ll T_{\rm w}$.

Extracting $m_a$ from the ratio $T_{\rm ann}/T_{\rm w}$ one gets,
\begin{equation}
    m_a \simeq 4.2 \times 10^{-2} \, \mathrm{GeV}~ \left(\frac{10^{-2}}{T_{\rm ann}/ T_{\rm w}}\right)^2~ \left( \frac{g_\star(T_{\rm w})}{105} \right)^{1/2} \left( \frac{T_{\rm ann}}{10^6 \, \rm{GeV}} \right)^2 ,
    \label{eqn:Tann/Tw ma bound}
\end{equation}
and equating this expression for $m_a$ with that in  Eq.~\eqref{eqn:Tw/V ma bound}, we obtain a relation which does not contain $m_a$ and can be written as 
\begin{equation}
   \frac{T_{\rm{ann}}}{10^6 \, \mathrm{GeV}} \simeq 8.9\, \frac{N^{1/3}}{f_\sigma^{1/6}} \left(\frac{105}{g_\star(T_{\rm w})}~\frac{105}{g_\star(T_{\rm{ann}})}\right)^{1/12}   \left( \frac{10^{-10} M_\odot}{M_{\rm PBH}}\right)^{1/3}\left(\frac{(T_{\rm{w}}/V)}{10^{-2}} \right)^{2/3} \left( \frac{(T_{\rm{ann}}/T_{\rm{w}})}{10^{-2}} \right)^{1/3}.
\end{equation}
Thus, requiring $T_{\rm{w}}/V <10^{-2}$ and $T_{\rm{ann}}/T_{\rm{w}} < 10^{-2}$, this last expression imposes an $M_{\rm PBH}$ dependent upper limit on $T_{\rm{ann}}$ shown as the blue thick line in Fig.~\ref{Fig Tann MPBH} and Fig.~\ref{Fig OmegaGW f}.

Finally, let us ensure that wall-domination never happens before annihilation. Had the walls not annihilated previously, the walls would dominate over radiation when $\sigma/t_{\rm wd} \simeq \rho(T_{\rm wd})$ or
\begin{equation}
 H(T_{\rm wd})= \frac{16 \pi}{3} \frac{\sqrt{2} f_\sigma}{N^2} \left(\frac{V}{M_{\rm P}}\right)^2 m_a\simeq 2.7 \times 10^{-6}   \left(\frac{V}{10^{16}~{\rm GeV}}\right)^2 m_a~.
\end{equation}
Thus, $H(T_{\rm wd}) < H(T_{\rm ann}) \simeq 0.33 \, (T_{\rm ann}/ T_{\rm w})^2\,  m_a$ means that
\begin{equation}
\label{Twd < Tann condition}
 \frac{V}{10^{16} \,{\rm GeV}} < 3.5 \times 10^2 \left(\frac{T_{\rm ann}}{T_{\rm w}}\right).
\end{equation}
This condition is automatically fulfilled for ${T_{\rm ann}}/{T_{\rm w}} = 10^{-2}$ since we require ${V}<10^{16}\,\rm GeV$. For smaller values of this ratio we check now that Eq.~\eqref{Twd < Tann condition} is fulfilled. In terms of temperatures this condition is
\begin{equation}
            \frac{T_{\rm wd}}{T_{\rm ann}} \simeq 4.8 \times 10^{-4}  \left(\frac{105}{g_\star (T_{\rm {ann}})} \right)^{1/4}\left(\frac{105}{g_\star (T_{\rm {wd}})} \right)^{1/4} \left( \frac{10^6 \, \rm{GeV}}{T_{\rm{ann}}} \right)^2 \left(\frac{10^{-10}  M_\odot}{M_{\rm{PBH}}} \right) < 1.
            \label{eqn:ratio4}
    \end{equation}
This limit rejects the region below and to the left of the dashed green line in Fig.~\ref{Fig Tann MPBH},  and thus does not constrain our regions of interest.

\section{Concluding remarks}

Many extensions of the Standard Model of particle physics feature $U(1)$ symmetries whose breaking is associated with the existence of heavy pGBs. There are QCD axion, ALP, and majoron models of this type. If the pGBs are unstable and decay in the early Universe, they cannot constitute the DM, and thus such models might lack a particle DM candidate.

Potentials which admit a number $N>1$ of minima along the orbit of vacua imply the formation of a string-wall network whose annihilation is accompanied by catastrogenesis~\cite{Gelmini:2022nim}, i.e. the production of GWs, pGBs, and PBHs. We find that for  annihilation temperatures $10^{4}\,{\rm GeV}\lesssim T_{\rm ann}\lesssim10^{9}\,\rm GeV$, one can form asteroid-mass PBHs ($10^{-16}\,M_\odot\lesssim M_{\rm PBH}\lesssim10^{-10}\, M_\odot$) that can constitute the entirety of the dark matter. For slightly smaller annihilation temperatures, $1\,{\rm GeV}\lesssim T_{\rm ann}\lesssim10^{5}\,\rm GeV$, the PBHs produced have mass $10^{-10}\,M_\odot\lesssim M_{\rm PBH}\lesssim10\,M_\odot$, and could potentially account for
HSC~\cite{Niikura:2017zjd} and other~\cite{EROS-2:2006ryy} microlensing observations, and they can account for part of the mergers observed by LIGO~\cite{Sasaki:2016jop}. The resulting PBH masses depend on $\alpha$, an exponent that parameterizes the annihilation rate of the string-wall network and needs to be determined from simulations of this process.

The GW signal produced by catastrogenesis together with PBHs can have an amplitude as large as $\Omega_{\rm GW}h^2\simeq 10^{-7}$ and can be observable in future GW detectors. For a particularly favorable set of parameters, the GWs emitted from strings alone, before walls appear, could be observable as well.

Notice that although in this paper we do not study the breaking of purely discrete symmetries, in principle our mechanism also applies to this case. The main ingredient is the existence of multiple vacua. Therefore, the breaking of a discrete symmetry can also result in a scenario similar to the one described here,  as far as the wall system annihilates mostly into unstable particles, such as flavons~\cite{Gelmini:2020bqg}, whose decay products thermalize.

There are several caveats concerning the interpretation of our results. The production of PBHs relies on the formation of closed domain walls which must retain a high degree of sphericity during the last stages of their collapse. 
The amplitude of the stochastic GW background produced by catastrogenesis is delicately sensitive to the precise evolution of the string-wall network at the very end of its existence.
Clarifying both these issues requires dedicated simulations of the process of annihilation of the network, which we hope will be tackled in the future.
The collapse of string-wall networks associated with heavy ALPs is an open problem worth examining in further detail.

\smallskip
\acknowledgments
{ The work of GG was supported in part by the U.S. Department of Energy (DOE) Grant No. DE-SC0009937. EV acknowledges support
by the European Research Council (ERC) under the European Union’s Horizon Europe
research and innovation programme (grant agreement No. 101040019). Views and opinions
expressed are however those of the author(s) only and do not necessarily reflect those of the
European Union.}

  \appendix

  \section{Appendix: Brief derivation of the present GW energy density}\label{sec:appendix}

The estimate of the power $P$ emitted in GWs is obtained from the quadrupole formula $P\simeq G\dddot{Q}_{ij}\dddot{Q}_{ij}$~\cite{Maggiore:1900zz}. The energy in the walls in the scaling regime is $E_{\rm w} \simeq \sigma t^2$, thus the quadrupole moment of the walls is $Q_{ij} \simeq~E_{\rm  w} t^2$ and  $\dddot{Q}_{ij} \simeq \sigma t$. Therefore  $P \simeq G \sigma^2 t^2$.  The energy density $\Delta \rho_{\rm GW}$ emitted by the string-wall system in a time interval $\Delta t$ is then 
$\Delta \rho_{\rm GW} (t) \simeq P \Delta t /t^3 \simeq G \sigma^2 {\Delta t}/{t}$. 
Thus, in a time interval equal to the Hubble time $\Delta t \simeq t$,   $\Delta \rho_{\rm GW} (t) \simeq G \sigma^2$ independently of
 the emission time $t$. This energy density is redshifted so its contribution to the present-day GW energy density is $\Delta \rho_{\rm GW} (t)(R(t)/R_0)^4 \simeq  G \sigma^2  (R(t)/R_0)^4$, where $R(t)$ and $R_0=1$ are the scale factors of the Universe at time $t$ and at present. 
 
 This argument clearly shows that the largest contribution to the present GW energy density spectrum emitted by the string-wall network, i.e. its peak amplitude, corresponds to the latest emission time,  $t \simeq t_{\rm ann}$, i.e.  
$\rho_{\rm GW}|_{\rm peak} \simeq G \sigma^2 (R(t_{\rm ann})/R_0)^4$.  Defining as usual
 $\Omega_{\rm GW}h^2|_{\rm peak} = \rho_{\rm GW}|_{\rm peak} (h^2/\rho_c)$, this leads to Eq.~\eqref{eq:OmegaGW-walls}.
Here  $\rho_c$ is the present critical density, $h$ is the reduced Hubble constant, and entropy conservation implies that $g_{s \star} (t) [R(t) T(t)]^3$ is constant, and at present the number of entropy degrees of freedom is $g_{s \star} (t_0)\simeq 3.93$~\cite{Saikawa:2018rcs}.

 What is usually quoted is the maximum of the GW energy density spectrum at time $t$ as a function of the wave-number at present $k$, which for $R_0=1$ coincides with the comoving wave-number (or the frequency $f=k/(2\pi)$), defined as
\begin{equation}
\Omega_{\rm GW}h^2 (k, t) = \left(\frac{h^2}{\rho_c(t)}\right)
 \left(\frac{d \rho_{\rm GW} (t)}{d \ln k}\right) ,
\end{equation}
(see e.g. Refs.~\cite{Maggiore:1900zz,Gelmini:2020bqg}).
This is also given by Eq.~\eqref{eq:OmegaGW-walls}, since 
\begin{equation} \label{d lnk}
\frac{d\rho_{\rm GW}(t)}{d\ln(k)}\simeq G \sigma^2
\end{equation}
independently of $t$. This is so because in the scaling regime the characteristic frequency of the GWs emitted at $t$ is $\simeq H(t)$ (the inverse of the horizon size). Thus the present-day frequency of GWs emitted at time $t$ is $f \simeq R(t) H(t)$. For GWs emitted in the radiation dominated epoch, when $H(t)=(2t)^{-1}$,  $d \ln f= (H(t)- t^{-1}) dt$, and  $d \ln f = d \ln k =-(1/2)~ d \ln t$.  Using  from above that
$d \rho_{\rm GW} (t) \simeq  G \sigma^2 (d t/  t)= G \sigma^2 d {\rm ln}(t)$, we find Eq.~\eqref{d lnk}.
 Thus the peak amplitude of this GW spectrum at present, for $t= t_0$, coincides with the result in Eq.~\eqref{eq:OmegaGW-walls}. Since the peak GW density of walls is emitted at annihilation, its present frequency is $f_{\rm peak} \simeq R(t_{\rm ann}) H(t_{\rm ann})$, as given in Eq.~\eqref{f-peak}.

The spectrum of the GWs emitted by  cosmic walls cannot be determined from the previous considerations alone, but has been computed numerically for $N>1$ in  Ref.~\cite{Hiramatsu:2012sc}. Figure~6 of Ref.~\cite{Hiramatsu:2012sc} shows that there is a peak at $k|_{\rm peak} \simeq R(t_f) m_a$, a bump at the scale $k\simeq R(t_f) H(t_f)$ where $t_f$ is the latest time in their simulation, and the spectral slope changes at these two scales. Causality requires that for frequencies below the peak, $k < k_{\rm peak}$ corresponding to  super-horizon wavelengths at $t_{\rm{ann}}$, 
the spectrum goes as $k^3$. This is a white noise spectrum characteristic of the absence of causal correlations~\cite{Caprini:2009fx}. At frequencies above the peak, $k > k_{\rm peak}$,  the spectrum depends instead on the particular GW production model assumed. The spectrum $1/k$ was found analytically for a source that is not correlated at different times, i.e. that consists of a series of short events~\cite{Caprini:2009fx} (see also Ref.~\cite{Cai:2019cdl}). Ref.~\cite{Hiramatsu:2012sc} obtained roughly a  $1/k$ spectrum for $k > k_{\rm peak}$,  with approximate slope and height of the secondary bump that depend on $N$.

 Similarly to what is done for the string-wall system, the  estimates of the GW density emitted by the string network before walls appear are based on the quadrupole formula (see e.g. Refs.~\cite{Chang:2019mza,Gouttenoire:2019kij,Gorghetto:2021fsn} and references therein). The energy of the string network is $E_{\rm st} \simeq \mu H^{-1}$, where $\mu$ is the string mass per unit length. Thus $\dddot{Q}_{ij} \simeq \mu$, and the power emitted in GWs is  $P \simeq G \mu^2$.  Using the same assumptions as for the walls, the energy density $\Delta \rho_{\rm GW}$ for strings is $\Delta \rho^{\rm{st}}_{\rm GW} (t)\simeq G \mu^2 (\Delta t)$.  The resulting spectrum has been computed in Refs.~\cite{Chang:2019mza,Gouttenoire:2019kij,Gorghetto:2021fsn}, and the simple expression in Eq.~\eqref{OmegaGW-strings} provides a good fit to all of them, for frequencies $f> f^{\rm st}_{\rm cut}$. The cutoff frequency $f^{\rm st}_{\rm cut}$ in Eq.~\eqref{fst-cut} corresponds to the frequency emitted at $T_{\rm w}$ when walls appear and the string network ceases to exist.

\bibliographystyle{bibi}
\bibliography{bibliography}
\end{document}